\documentclass[sigconf,authorversion]{acmart}

\usepackage{enumitem}
\usepackage{listings}
\usepackage{subcaption}

\graphicspath{{./figures/}}
\DeclareGraphicsExtensions{.pdf,.jpeg,.png}

\copyrightyear{2023} 
\acmYear{2023} 
\setcopyright{acmlicensed}\acmConference[EASE '23]{Proceedings of the International Conference on Evaluation and Assessment in Software Engineering}{June 14--16, 2023}{Oulu, Finland}
\acmBooktitle{Proceedings of the International Conference on Evaluation and Assessment in Software Engineering (EASE '23), June 14--16, 2023, Oulu, Finland}
\acmPrice{15.00}
\acmDOI{10.1145/3593434.3593501}
\acmISBN{979-8-4007-0044-6/23/06}

\begin{document}

\title{Outside the Sandbox: A Study of Input/Output Methods in Java}

\author{Mat\'u\v{s} Sul\'ir}
\email{matus.sulir@tuke.sk}
\orcid{0000-0003-2221-9225}
\affiliation{%
  \institution{Technical University of Ko\v{s}ice}
  \streetaddress{Letn\'a 9}
  \postcode{04200}
  \city{Ko\v{s}ice}
  \country{Slovakia}
}

\author{Sergej Chodarev}
\email{sergej.chodarev@tuke.sk}
\orcid{0000-0002-9293-0859}
\affiliation{%
  \institution{Technical University of Ko\v{s}ice}
  \streetaddress{Letn\'a 9}
  \postcode{04200}
  \city{Ko\v{s}ice}
  \country{Slovakia}
}

\author{Milan Nos\'a\v{l}}
\email{milan.nosal@gmail.com}
\orcid{0000-0002-8831-9440} % Is this correct?
\affiliation{%
  \institution{ValeSoft, s.r.o.}
  \streetaddress{Vod\'arensk\'a 636/3}
  \postcode{04001}
  \city{Ko\v{s}ice}
  \country{Slovakia}
}

\renewcommand{\shortauthors}{Sul\'ir et al.}

\begin{abstract}
Programming languages often demarcate the internal sandbox, consisting of entities such as objects and variables, from the outside world, e.g., files or network. Although communication with the external world poses fundamental challenges for live programming, reversible debugging, testing, and program analysis in general, studies about this phenomenon are rare. In this paper, we present a preliminary empirical study about the prevalence of input/output (I/O) method usage in Java. We manually categorized 1435 native methods in a Java Standard Edition distribution into non-I/O and I/O-related methods, which were further classified into areas such as desktop or file-related ones. According to the static analysis of a call graph for 798 projects, about 57\% of methods potentially call I/O natives. The results of dynamic analysis on 16 benchmarks showed that 21\% of the executed methods directly or indirectly called an I/O native. We conclude that neglecting I/O is not a viable option for tool designers and suggest the integration of I/O-related metadata with source code to facilitate their querying.
\end{abstract}




\maketitle

%%%%%%%%%%%%%%%%%%%%%%%%%%%%%%%%%%%%%%%%%%%%%%%%%%%%%%%%%%%%

\section{Introduction}

In many contemporary programming languages, programs are executed inside a virtual machine that separates two worlds: the internal sandbox and the external environment. The internal sandbox consists of executable code and data, such as classes, objects and their fields on the heap, local variables on the stack, etc. The external environment is accessible by calling input/output (I/O) methods designated for file manipulation, inter-process communication, and the utilization of other operating system services. In imperative programming languages such as Java, non-I/O and I/O constructs can interleave without restrictions. This simplifies programming from the cognitive viewpoint, but it has far-reaching consequences for the design of advanced development tools.

As a motivational example, consider the live programming environment SEEDE by Reiss et al. \cite{Reiss18seede}. First, the developer chooses a method of interest and places a special breakpoint in it. When the program reaches the beginning of this method during execution in a debugger, a snapshot of the complete Java Virtual Machine (JVM) process at this point is created. Now the developer can modify the source code of the method and re-run it, using the saved snapshot as an entry point. The modification and re-running of the method can be repeated multiple times (manually or potentially after every keystroke), each time offering a view of the method's effects on the program state in the debugger.

As long as the method of interest contains only non-I/O constructs, the tool works perfectly. Consider this example:
\begin{lstlisting}
public void computeHeader() {
  byte[] header = {0, 0, 0};
  if (flag)
    header[0] = 1 << 4;
  ...
}
\end{lstlisting}

Then the developer decides to add an I/O method call:
\begin{lstlisting}
public void computeHeader() {
  byte[] header = {0, 0, 0};
  if (flag)
    header[0] = 1 << 4;
  ...
  fileStream.write(header);
}
\end{lstlisting}

From this moment, using the live programming approach can have unintended consequences. Re-running the method multiple times writes duplicate headers to the file, resulting in corrupted data.

The best way to deal with this problem is to design special handlers for every I/O call. For instance, standard file manipulation routines could be intercepted by writing all changes into a temporary overlay file system and discarding them after each reset of the internal program state. The creation of such handlers is an expensive task though. Furthermore, some I/O operations cannot be practically intercepted and reversed in general: consider an HTTP POST request that modifies data on a third-party server.

Therefore, we might ponder how widespread is this problem. \textit{Would it not be possible to simply disregard it?} If a vast majority of Java methods did not invoke any I/O calls, live programming would be usable to a great extent even without special handlers.

Although we explained the issue using a live programming example, it is certainly not limited to it. Analogous fundamental problems occur in reversible debugging \cite{Pothier09back}, software testing (e.g., automated test generation \cite{Arcuri14automated} and mocking), and many other areas.

Despite their importance, I/O methods themselves have rarely been studied empirically so far. Rocha et al. \cite{Rocha19compehending} analyzed the energy behavior of a small number of selected Java I/O APIs. Grichi et al. \cite{Grichi19state} qualitatively analyzed usage practices of native methods, i.e., Java Native Interface, without specifically considering I/O methods, which represent only their subset. Figueroa et al. \cite{Figueroa21which} studied the prevalence of monads, including the \texttt{IO} monad, in the functional programming language Haskell.

To the best of our knowledge, this is the first study on the prevalence of I/O methods in an imperative language. Therefore, we would like to answer the following research questions in this paper:

\begin{description}[leftmargin=0cm]
\item[RQ1] How can native methods present in the Java Runtime Environment (JRE) be categorized with respect to their I/O effects?
\item[RQ2] What portion of methods in Java projects potentially call I/O natives according to static analysis?
\item[RQ3] What portion of methods executed in a sample of Java benchmarks actually call I/O natives?
\end{description}

\section{Classification of Native Methods}

In Java, the only way to figuratively break the sandbox and communicate with the outside world (I/O) is through native methods -- methods marked with the \texttt{native} keyword and implemented in C or C++. On the other hand, not all native methods represent I/O actions. Some of them are native merely because of performance optimization or to provide advanced reflective operations. To answer RQ1, we manually categorized native methods into I/O and non-I/O ones.

\subsection{Method}

We focused only on natives contained directly in the Java Development Kit (JDK), not developer-written native methods packaged as dynamic libraries with specific projects. As a reference implementation, we chose the latest available long-term release, Eclipse Temurin 17 for Linux\footnote{jdk-17.0.6+10, Linux x64 from \url{https://adoptium.net/temurin/releases/}}. From this distribution, we created a custom JRE containing only modules named \texttt{java.*}, i.e., modules belonging to Java Standard Edition (SE) Platform plus \texttt{java.smartcardio}. This JRE was used throughout the rest of the study.

From the mentioned JRE, we extracted all native methods, including the private and package-private ones, using bytecode analysis. This resulted in a list of 1435 methods across 8 modules.

Instead of merely marking the methods as I/O or non-I/O, we decided to divide them into more specific categories, such as desktop-related or filesystem operations. To construct a set of categories and agree on their definitions, we first performed a pilot study on a small subset of methods. The first author tagged each of these methods using open qualitative coding \cite{Saldana16coding}, i.e., without having a fixed set of codes. He gradually merged and modified them to a set of fixed categories, with each method pertaining only to one category. The second and third authors tagged the methods by closed coding, using this preliminary set of possible categories, while also noting possible ambiguities. Then all three authors mutually resolved their disputes and agreed on the set of categories and their definitions.

Next, all 1435 methods were categorized by the first author. At the same time, the second and third authors classified a random subset of 25\% and 75\% methods, respectively. Each method was therefore independently categorized by two different researchers.

Finally, all three researchers met and resolved their disputes by mutual agreement. At this meeting, minor changes to the taxonomy itself were also performed, such as merging two similar categories.

During the whole classification process, all researchers used the following guidelines. First, they considered the name of the corresponding class and method together with their documentation. Next, they searched for their callers in the Java code to get an intuition about their usage. Finally, they inspected the C or C++ code implementing the given method. Logging statements, such as assertions, were excluded from the analysis since they are disabled by default. If the method could pertain to more than one I/O category, we chose the more specific one, e.g., reading a desktop environment configuration file was categorized as ``desktop'' instead of ``files''.

\subsection{Results}
\label{s:classify-results}

Table~\ref{t:categories} displays the final set of categories of native Java methods, along with their descriptions and the counts of corresponding methods in the analyzed JRE.

\begin{table*}
\caption{The Categories of Native Methods}
\label{t:categories}
\centering
\begin{tabular}{lp{14.5cm}r}
\toprule
Category & Description & Count \\
\midrule
non-io & Code that bridges the pure Java code with the native space inside the current JVM process. This includes optimized algorithm implementations, reflection, classloading, threading, native data structure initialization, no-ops, etc. & 623\\
invocation & An operation representing an arbitrary reflective method invocation. Such a method could be I/O or non-I/O, but we cannot determine this with static analysis in general. & 17\\
desktop & Desktop-related operations: drawing to screen or graphics card’s buffers, manipulating windows and widgets, processing keyboard and mouse input, playing sounds, printing, etc. & 416\\
time & Operations related to current system time, time measurement, and time zones. & 28\\
files & Operations on standard files, directories, and file systems. & 135\\
network & Operations for communication with other processes via network, sockets, or pipes and methods related to network adapters. & 111\\
os & Miscellaneous operating system calls, such as methods related to resource limits, security, processes, environment variables, shared memory, etc. & 105\\
\bottomrule
\end{tabular}
\end{table*}

The first category, ``non-io'', contains methods not representing communication with the outside world. Except for classic cases such as algorithmic optimization or reflection, we also included, perhaps controversially, threading and classloading in this category.

Although threading is often implemented using operating system threads, the JVM Specification does not prescribe this. Furthermore, threads are omnipresent in Java: every thread can be paused and resumed at arbitrary times, which leads to situations such as race conditions. These are the subject of a multitude of studies and are out of our scope. Finally, we suppose the designers of tools such as live programming environments will pose restrictions on the threading behavior, e.g., enable the debugging of only one thread while the rest are suspended.

Classloading can be, and often is, performed lazily in Java. This means it is highly non-deterministic: calling an innocent-looking method can invoke a custom classloader that performs a network request to load a given class since it is used for the first time during the application execution. We suppose tool authors will implement an eager classloader loading classes only from a fixed set of local JAR files -- which we consider to be a part of the Java's sandbox.

The second category in Table~\ref{t:categories}, ``invocation'', consists of natives whose purpose is to call an arbitrary user-supplied method. Such a method could be an I/O or non-I/O one, and we cannot determine this beforehand. To maximize soundness, we consider the ``invocation'' category to be part of the I/O natives during static analysis. During dynamic analysis, we consider it non-I/O because the captured call stack includes information about specific native methods called using reflection.

The remaining five categories represent various I/O-related calls: ``desktop'', ``time'', ``files'', ``network'', and ``os'' (miscellaneous operating system operations). Each category has certain specifics from the viewpoint of tool developers, especially of live programming environments. For instance, system time could be simulated if necessary. File operations could be possibly replaced by an overlay file system. On the other hand, simulating network calls is practically impossible in general, as we already mentioned.

\section{Static Analysis of I/O Calls}

In this section, we will determine what portion of methods in Java projects can directly or indirectly call native JRE methods, especially the I/O ones, to answer RQ2.

\subsection{Method}

The method consists of two parts: the construction of a corpus of Java projects and their subsequent static analysis.

\subsubsection{Dataset}

Our aim is to statically analyze the call graphs of Java projects down to the lowest layers of the JDK, so we need their binary form, including complete dependencies. Since we chose a recent JDK version in RQ1, the dataset has to be as up-to-date as possible due to the incompatibilities introduces by new Java releases \cite{Sulir20large}. Furthermore, we have special requirements on selected properties of the dataset, e.g., it cannot contain non-JRE native methods. Therefore, we opted for the creation of our own corpus.

We downloaded and built a set of open-source projects from GitHub, using the following inclusion criteria:
\begin{description}[leftmargin=0cm]
\item[I1] The main programming language of the project is Java.
\item[I2] A machine-recognizable open-source license file is contained in the repository.
\item[I3] The repository has at least 10 stars. This is a limitation imposed by the used search tool. At the same time, it can be considered a high-precision strategy to include only engineered software projects and exclude toy or sample projects and homework assignments \cite{Munaiah17curating}.
\item[I4] The repository contains an automated build configuration file in its root directory. We focused on projects using Maven (pom.xml) since it is the most popular build system for Java \cite{Sulir20large}, and it offers a consistent way to retrieve the built artifacts, including all transitive dependencies.
\item[I5] The project can be successfully built from source code using the supplied configuration file on JDK 17 without manual intervention.
\item[I6] The set of produced JAR files contains at least one entry point, namely a \texttt{main} method or a JUnit (versions 3 to 5) test method. This was necessary since call graph construction algorithms require the specification of entry points to function properly.
\end{description}

We further refined the set of projects using the following exclusion criteria:
\begin{description}[leftmargin=0cm]
\item[E1] Forks.
\item[E2] Duplicates (based on the target JAR file names) that are not explicitly marked as forks.
\item[E3] Projects containing unresolved references to other classes in the resulting binary files. This is necessary to construct call graphs that are sound to the extent practically possible.
\item[E4] Projects using Java Native Interface (JNI) via custom developer-written methods. Since only methods included in the JRE were manually categorized, we could not distinguish between the I/O and non-I/O custom natives.
\item[E5] Projects utilizing JDK modules not included in our custom JRE. Again, such modules could contain native methods that were not categorized.
\end{description}

To construct the dataset, we first obtained a list of all repositories matching criteria I1--I3 and E1 using the GitHub Search tool by Dabic et al. \cite{Dabic21sampling}. We downloaded the source code of repositories matching I4 with GitHub API. Then we tried to automatically build them (I5) using Maven in a Docker container having the necessary tools installed. We checked criterion E2 with a shell script. The JAR files including dependencies were checked for criteria I6 and E4 using the \texttt{javap} disassembler, and criteria E3 and E5 with the dependency analysis tool \texttt{jdeps}.

This process resulted in a list of 807 projects. Subsequent call graph generation failed for 9 projects due to technical reasons. The final dataset thus consists of 798 Java projects from diverse domains and of various sizes, ranging from 2 to 14,621 methods per project (not counting dependencies).

\subsubsection{Static Analysis}

As a next step, we performed static analysis on the created corpus. We define the \textit{source method} as a Java method that is a part of the project itself (not the dependencies), is not synthetic (generated by the compiler and missing in the original source code), and is not a static initializer. We excluded class initializers since we assume classloading is already finished before the execution of each method, as we discussed in section \ref{s:classify-results}. The total number of source methods in the corpus was 550,727.

For every project, we constructed a whole-program static call graph including its dependencies and our complete JRE. We used the SPARK algorithm in the Soot library \cite{Vallee-rai10soot}, which offers reasonable precision without sacrificing scalability.

We define a \textit{reachable source method} as an entry point (a \texttt{main} method or a test) or a source method that can be directly or indirectly called from at least one entry point. About 57.2\% of the source methods in our dataset were reachable.

For every reachable source method (sometimes abbreviated as a caller in the following text), we determined a set of native methods (natives) that can be directly or indirectly called from it. These sets along with associated metadata were stored in a relational database for further querying.

\subsection{Results}

From all reachable source methods, 62.6\% can potentially call a native method and 56.9\% can call an I/O native method according to the static analysis. This means I/O-involving methods are very frequent, and neglecting them during any tool design is unacceptable.

In Table~\ref{t:categories-static}, there is a list of the categories of natives along with the percentages of reachable source methods that can potentially call at least one native method from a given category. Non-I/O native methods are the most frequent. Natives from five categories (time, files, os, network, invocation) follow closely behind each other. This means the I/O native methods in the JRE are highly interconnected, and as soon as one of them is called, it is difficult to distinguish with static analysis which kind of I/O call will be made. Finally, desktop I/O methods are the least frequent ones since many applications and libraries do not support or handle graphical user interfaces.

\begin{table}
\caption{The portions of analyzed/executed methods calling native methods from a given category.}

\begin{subtable}{0.45\linewidth}
\caption{static analysis}
\label{t:categories-static}
\centering
\begin{tabular}{lr}
\toprule
Category & Portion \\
\midrule
non-io & 62.44\% \\
invocation & 56.71\% \\
desktop & 23.07\% \\
time & 56.91\% \\
files & 56.74\% \\
network & 56.71\% \\
os & 56.72\% \\
\bottomrule
\end{tabular}
\end{subtable}
\hfill
\begin{subtable}{0.45\linewidth}
\caption{dynamic analysis}
\label{t:categories-dynamic}
\centering
\begin{tabular}{lr}
\toprule
Category & Portion \\
\midrule
non-io & 41.29\% \\
invocation & 6.70\% \\
desktop & 0.11\% \\
time & 18.75\% \\
files & 18.69\% \\
network & 0.78\% \\
os & 1.76\% \\
\bottomrule
\end{tabular}
\end{subtable}

\end{table}

\section{Dynamic Analysis of I/O Calls}

To answer RQ3, we will describe the method and results of dynamic analysis on a suite of Java benchmarks.

\subsection{Method}

As a basis to construct the corpus used for dynamic analysis, we selected the DaCapo \cite{Blackburn06dacapo} benchmark suite\footnote{commit \href{https://github.com/dacapobench/dacapobench/tree/04132797de831d3e2afaf05c03a3dfdfe3991e8d}{0413279}}. It provides 22 nontrivial, real-life workloads from various domains. First, we excluded 6 benchmarks incompatible with our JRE. Next, to make dynamic analysis feasible despite the dramatic slowdown it imposes, we filtered the associated data for each benchmark to include only the ``small'' size. The resulting set of 16 benchmarks is available online\footnote{\url{https://osf.io/2jqpv}}.

We executed each benchmark in a debugging mode and traced all method entry events. Every time a method was called, we added it to a set of executed methods unless it was a static initializer, a JRE built-in, synthetic, or abstract method. Whenever a native method was called, we obtained the current call stack and marked all methods present in it as its callers.

Five benchmarks failed due to timeouts but we included them in the results since they represent valid real-life workloads too. The results were saved to a database for querying.

\subsection{Results}

From all non-excluded executed methods, 41.9\% called a native method and 21.4\% called an I/O native. This confirms that I/O methods are frequently called on actual workloads.

The portions of executed native methods for each category are shown in Table~\ref{t:categories-dynamic}. The most frequent categories are ``time'' and ``files'', which is understandable since the synopsis of many benchmarks is to load files and measure the time of their processing.

\section{Discussion}

Now we will take a closer look at the results and discuss possible reasons and implications.

\begin{figure}
\includegraphics[width=\linewidth]{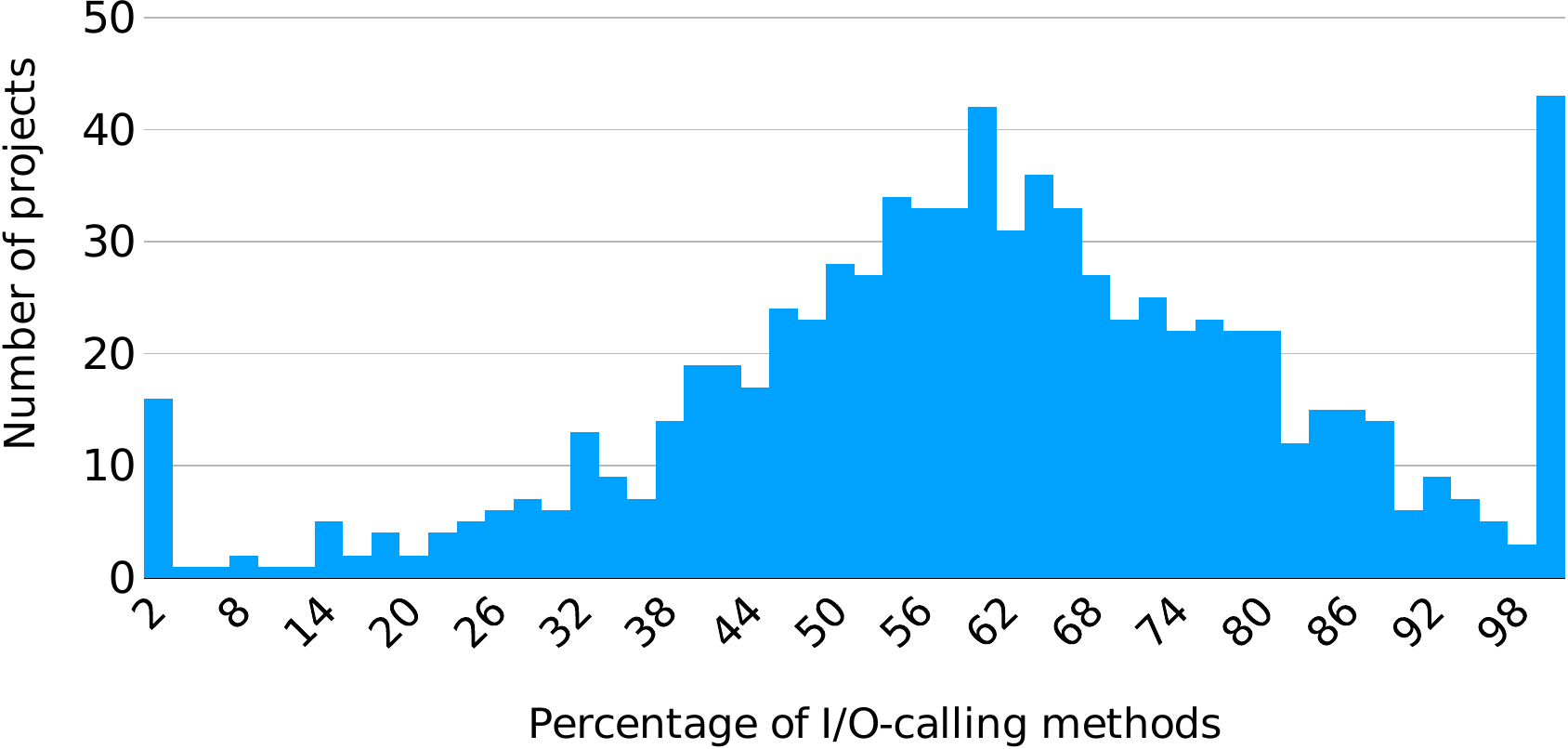}
\caption{The distribution of I/O callers percentage per project (static analysis)}
\label{f:project}
\end{figure}

First, from the project-based view, we might ask whether there are differences in the proportion of I/O callers among projects. For example, a mathematical library should call I/O natives only scarcely, while in a network monitor, they could be almost omnipresent. Figure~\ref{f:project} displays a distribution of the percentages of I/O-calling methods for individual projects, determined using static analysis. The portions are approximately normally distributed, with the exception of the left and right extremes. The left extreme is represented mainly by collections of algorithms and projects for which the static analysis was unable to detect I/O calls despite their presence. The right extreme contains various smaller utilities, ranging from test framework extensions to simple data format parsers, often with a low number of reachable methods.

The dynamic analysis results of the project-based view were more uniform. The portion of I/O callers ranged from 17\% of the executed methods for H2 (an SQL database) to 36\% for GraphChi (a disk-based graph computation system). Twelve of the 16 benchmarks had the I/O callers portion between 18 and 25\%.

\begin{figure}
\includegraphics[width=\linewidth]{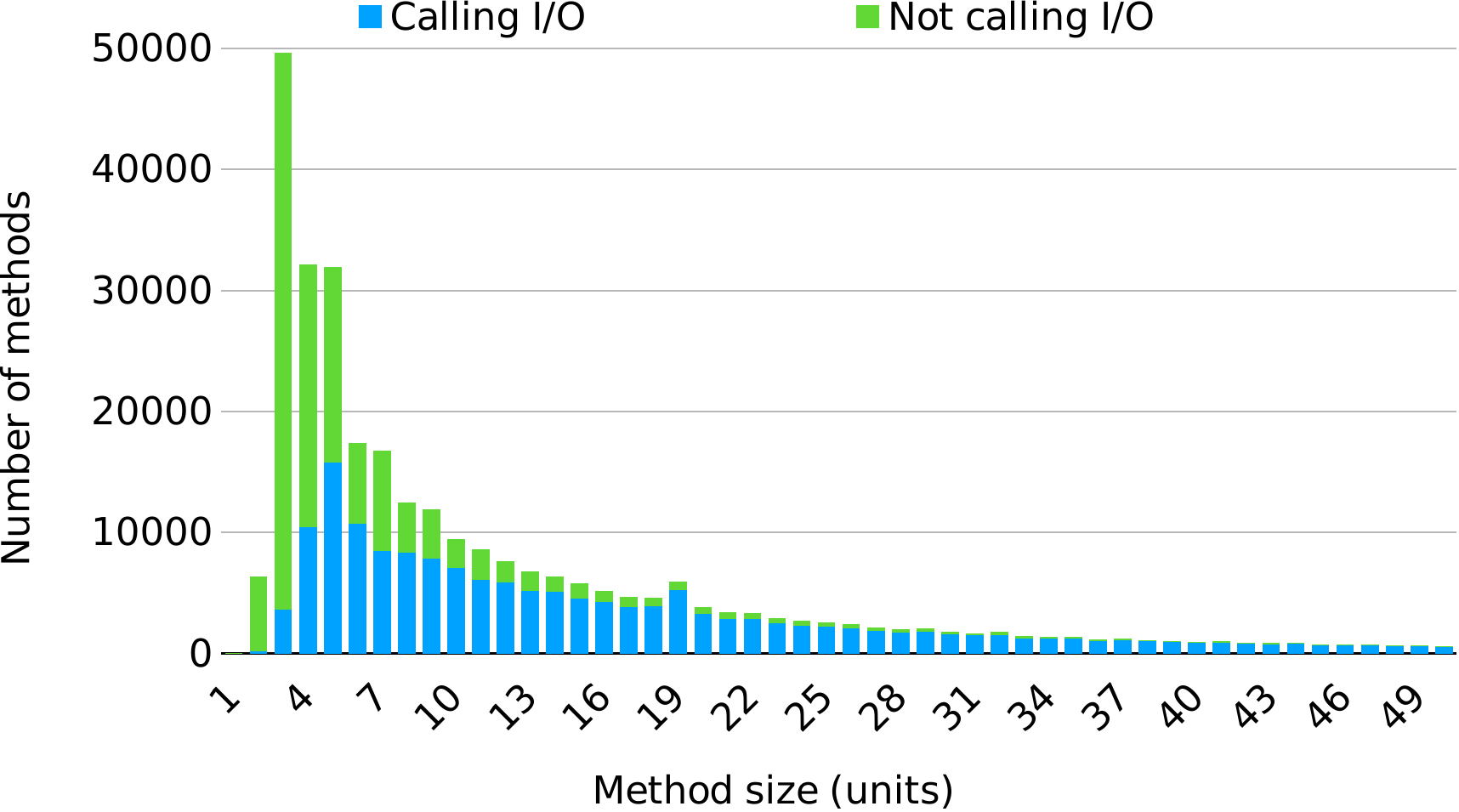}
\caption{The number of reachable source methods calling and not calling I/O per method size (static analysis)}
\label{f:size}
\end{figure}

Next, we would like to look at the relationship between the size of a method and the likelihood that it is a potential I/O caller. In Figure~\ref{f:size}, there are method sizes on the x-axis, measured in abstract units, where one unit represents a statement such as a variable assignment or a method call. The right side is cropped due to larger method sizes occurring rarely. On the y-axis are numbers of methods that potentially call or do not call I/O natives, as determined by static analysis. We can see that extremely short methods are usually non-I/O ones, while moderately-sized methods call I/O natives more often.

Taking into account only methods with at least five units, which we might consider a threshold for a non-trivial method worth debugging with advanced tools, 72.8\% of them potentially call I/O natives. Similarly, according to the dynamic analysis, 27.8\% of executed methods larger than 10 bytes (of bytecode) called I/O natives. This further confirms the fact that neglecting I/O is not an option.

Finally, we inspected a list of the specific most frequently called native methods per category in our classification. Many groups of methods had the same static caller counts, which confirmed our intuition that distinguishing which specific I/O call will be performed at runtime with static analysis is difficult in general. A typical example is a blurry line between files and network in Linux since both of them are represented by file descriptors in the kernel, which is reflected also in the JDK internals. Special files such as Unix domain sockets and pipes make the distinction even less clear.

For dynamic analysis, the differences in caller counts varied much more. The most frequent I/O methods were related to obtaining precise system time, along with seeking and reading files. Of course, the results are workload-specific.

\section{Threats to Validity}

\paragraph{Construct Validity}

Static analysis is inherently imprecise, and the portion of methods that actually invoke an I/O action during a particular execution is typically lower than the number of methods that potentially call them, as we showed. However, in Java, static analysis is also unsound. We observed two main reasons in our study. First, the analyzer can miss non-trivial reflective calls. We coped with this issue by creating a special category, ``invocation'', that represents executions of arbitrary methods via reflection, and counting it as a potential I/O call. Second, if a final member variable is assigned the value of \texttt{null} and then overwritten in native code (which is possible), the static analyzer does not register any method calls on this field. Solving this problem in general is difficult since native code is opaque to a Java static analyzer. The mentioned problem concerns variables such as \texttt{System.out}, providing standard I/O stream facilities. The impact of this issue on the results should be studied in future work.

\paragraph{Internal Validity}

During the creation of the static analysis dataset, we used multiple inclusion and exclusion criteria that limited the set of analyzed repositories to a relatively small number of projects with restricted properties. However, we provided a reasonable rationale for each criterion. One of the criteria that shrank the project count the most was the ability to be automatically built with JDK 17. This was often caused by the incompatibility of older projects with newer JDK versions. While we could focus on an older release, we opted for the latest long-term support version since a lot of manual effort was put into categorization, which could soon become obsolete in such a case.

\paragraph{External Validity}

The study was limited to Java, which marks methods allowed to directly perform I/O with its \texttt{native} keyword. Although probably less elegant, similar studies could be performed also in other languages.

All parts of the study were limited not only to a specific JDK version but also to an operating system, as a part of native methods is platform-specific. Although a similar study could be performed, e.g., for a Windows version of JDK 20, we would not expect dramatically different results.

The results of the dynamic analysis highly depend on the fact that benchmarks were selected as the studied programs. However, this could be also true, e.g., for unit tests since they often mock I/O. Ideally, we should execute a diverse set of automated workloads ranging from mathematical computation to user interface manipulation, which is difficult to obtain.

\paragraph{Reliability}

The categorization of the list of native JRE methods was performed manually, which could cause subjectivity in the results. However, we performed a pilot study to reveal the differences in the reasoning of individual researchers and provided clear definitions of each category. Most importantly, each native method was categorized independently by two researchers, and the final decision was performed by mutual agreement.

The automated part of our study is fully reproducible. We published the full source code at \url{https://github.com/sulir/iostudy} and the data in a permanent repository at \url{https://doi.org/10.17605/OSF.IO/CNSRJ}.

\section{Related Work}

A study by Rocha et al. \cite{Rocha19compehending} focused on a small number of I/O methods, namely 22 methods for reading and writing to streams. They analyzed their energy behavior and tried to improve it using a set of micro-benchmarks. Figueroa et al. \cite{Figueroa21which} studied the usage of monads in Haskell. It is, however, a purely functional programming language that strictly separates I/O and non-I/O code.

A study by Grichi et al. \cite{Grichi19state} offers a view into usage practices of Java Native Interface. By manual analysis of 100 open-source projects, they constructed a list of common patterns that developers utilize when connecting Java code with the native C/C++ implementation. They did not consider I/O methods in any way.

Mastrangelo et al. \cite{Mastrangelo15use} focused on the Java Unsafe API, which is implemented using native methods and offers to bypass Java's security mechanisms when accessing memory. The authors studied the prevalence of this API and its usage practices. Similarly, Tan and Croft \cite{Tan08empirical} analyzed the usage of native code in JDK from a security perspective, without considering I/O methods specifically.

The authors of ChroniclerJ \cite{Bell13chronicler}, a reversible debugger, also faced an issue that not all native methods in Java represent I/O calls, which they solved by manual annotation too. However, their goal was broader since they needed to distinguish between deterministic and non-deterministic methods in general. They also analyzed the list of public API methods potentially connected with the (often private) native methods, so their list was much longer, which also resulted in its incompleteness. Finally, the JDK version they analyzed is no longer supported.

\section{Conclusion}

In this paper, we presented a preliminary empirical study on the prevalence of I/O calls in Java projects. We manually divided a list of native Java methods in a selected JRE into seven categories, some of which we considered I/O. According to the static analysis results on a corpus of Java projects, 57\% of reachable source methods potentially communicate with the world outside the sandbox using I/O. During a dynamic analysis on a benchmark suite, 21\% of executed methods actually called I/O. These numbers are even higher when considering only non-trivial methods or when taking into account the potential unsoundness of static analysis.

Discerning natives that potentially perform I/O operations from natives operating inside the virtual machine is a common problem. The list of the categorized JRE methods is thus available as a part of the dataset at \url{https://doi.org/10.17605/OSF.IO/CNSRJ}. 

In many software engineering sub-areas, but particularly in the live programming community \cite{Rein18exploratory}, I/O is rarely taken into account when designing tools (with very few exceptions, such as the mentioned SEEDE \cite{Reiss18seede}). Therefore, we conclude tool authors cannot neglect I/O actions since they are almost omnipresent in programming languages such as Java.

Statically analyzing the I/O effects using whole-program analysis is difficult due to the interconnection of methods that implement them. Therefore, we propose the creation of special annotations for marking methods potentially performing input/output, e.g., \texttt{@IO} (alternatively, marking the rest of them \texttt{@NonIO}). Currently, analogous annotations exist for nullable/non-null values, pure methods that do not modify the state of objects, and many other contracts.

As a final motivation, having a clear distinction between I/O or non-I/O methods, e.g., with the help of the mentioned annotations, could make program analysis faster and more precise. For example, unit test generators would more easily know which parts of the system to mock. We could also query the development environment if the given method is non-I/O and can be safely used for live programming, especially if dangerous operations such as file deletion were noted using corresponding annotation parameters.

\begin{acks}
This work was supported by project VEGA No. 1/0630/22 Lowering Programmers' Cognitive Load Using Context-Dependent Dialogs.
\end{acks}

\bibliographystyle{ACM-Reference-Format}
\bibliography{ease}


\begin{thebibliography}{16}

%%% the Web address.

\ifx \showCODEN    \undefined \def \showCODEN     #1{\unskip}     \fi
\ifx \showDOI      \undefined \def \showDOI       #1{#1}\fi
\ifx \showISBNx    \undefined \def \showISBNx     #1{\unskip}     \fi
\ifx \showISBNxiii \undefined \def \showISBNxiii  #1{\unskip}     \fi
\ifx \showISSN     \undefined \def \showISSN      #1{\unskip}     \fi
\ifx \showLCCN     \undefined \def \showLCCN      #1{\unskip}     \fi
\ifx \shownote     \undefined \def \shownote      #1{#1}          \fi
\ifx \showarticletitle \undefined \def \showarticletitle #1{#1}   \fi
\ifx \showURL      \undefined \def \showURL       {\relax}        \fi
% invisible to TeX
\providecommand\bibfield[2]{#2}
\providecommand\bibinfo[2]{#2}
\providecommand\natexlab[1]{#1}
\providecommand\showeprint[2][]{arXiv:#2}

\bibitem[Arcuri et~al\mbox{.}(2014)]%
        {Arcuri14automated}
\bibfield{author}{\bibinfo{person}{Andrea Arcuri}, \bibinfo{person}{Gordon
  Fraser}, {and} \bibinfo{person}{Juan~Pablo Galeotti}.}
  \bibinfo{year}{2014}\natexlab{}.
\newblock \showarticletitle{Automated Unit Test Generation for Classes with
  Environment Dependencies}. In \bibinfo{booktitle}{\emph{Proceedings of the
  29th ACM/IEEE International Conference on Automated Software Engineering}}
  \emph{(\bibinfo{series}{ASE '14})}. \bibinfo{publisher}{ACM},
  \bibinfo{address}{New York, NY, USA}, \bibinfo{pages}{79--90}.
\newblock
\showISBNx{9781450330138}
\urldef\tempurl%
\url{https://doi.org/10.1145/2642937.2642986}
\showDOI{\tempurl}


\bibitem[Bell et~al\mbox{.}(2013)]%
        {Bell13chronicler}
\bibfield{author}{\bibinfo{person}{Jonathan Bell}, \bibinfo{person}{Nikhil
  Sarda}, {and} \bibinfo{person}{Gail Kaiser}.}
  \bibinfo{year}{2013}\natexlab{}.
\newblock \showarticletitle{Chronicler: Lightweight Recording to Reproduce
  Field Failures}. In \bibinfo{booktitle}{\emph{Proceedings of the 2013
  International Conference on Software Engineering}}
  \emph{(\bibinfo{series}{ICSE '13})}. \bibinfo{publisher}{IEEE Press},
  \bibinfo{address}{Piscataway, NJ, USA}, \bibinfo{pages}{362--371}.
\newblock
\showISBNx{978-1-4673-3076-3}
\urldef\tempurl%
\url{https://doi.org/10.1109/ICSE.2013.6606582}
\showDOI{\tempurl}


\bibitem[Blackburn et~al\mbox{.}(2006)]%
        {Blackburn06dacapo}
\bibfield{author}{\bibinfo{person}{Stephen~M. Blackburn} {et~al\mbox{.}}}
  \bibinfo{year}{2006}\natexlab{}.
\newblock \showarticletitle{The {D}a{C}apo Benchmarks: {J}ava Benchmarking
  Development and Analysis}. In \bibinfo{booktitle}{\emph{Proceedings of the
  21st Annual ACM SIGPLAN Conference on Object-oriented Programming Systems,
  Languages, and Applications}}. \bibinfo{publisher}{ACM},
  \bibinfo{address}{New York, NY, USA}, \bibinfo{pages}{169--190}.
\newblock
\urldef\tempurl%
\url{https://doi.org/10.1145/1167473.1167488}
\showDOI{\tempurl}


\bibitem[Dabic et~al\mbox{.}(2021)]%
        {Dabic21sampling}
\bibfield{author}{\bibinfo{person}{Ozren Dabic}, \bibinfo{person}{Emad
  Aghajani}, {and} \bibinfo{person}{Gabriele Bavota}.}
  \bibinfo{year}{2021}\natexlab{}.
\newblock \showarticletitle{Sampling Projects in {G}it{H}ub for {MSR} Studies}.
  In \bibinfo{booktitle}{\emph{2021 IEEE/ACM 18th International Conference on
  Mining Software Repositories (MSR)}}. \bibinfo{pages}{560--564}.
\newblock
\urldef\tempurl%
\url{https://doi.org/10.1109/MSR52588.2021.00074}
\showDOI{\tempurl}


\bibitem[Figueroa et~al\mbox{.}(2021)]%
        {Figueroa21which}
\bibfield{author}{\bibinfo{person}{Ismael Figueroa}, \bibinfo{person}{Paul
  Leger}, {and} \bibinfo{person}{Hiroaki Fukuda}.}
  \bibinfo{year}{2021}\natexlab{}.
\newblock \showarticletitle{Which monads {H}askell developers use: An
  exploratory study}.
\newblock \bibinfo{journal}{\emph{Science of Computer Programming}}
  \bibinfo{volume}{201} (\bibinfo{year}{2021}), \bibinfo{pages}{102523}.
\newblock
\showISSN{0167-6423}
\urldef\tempurl%
\url{https://doi.org/10.1016/j.scico.2020.102523}
\showDOI{\tempurl}


\bibitem[Grichi et~al\mbox{.}(2019)]%
        {Grichi19state}
\bibfield{author}{\bibinfo{person}{Manel Grichi}, \bibinfo{person}{Mouna
  Abidi}, \bibinfo{person}{Yann-Ga\"{e}l Gu\'{e}h\'{e}neuc}, {and}
  \bibinfo{person}{Foutse Khomh}.} \bibinfo{year}{2019}\natexlab{}.
\newblock \showarticletitle{State of Practices of {J}ava {N}ative {I}nterface}.
  In \bibinfo{booktitle}{\emph{Proceedings of the 29th Annual International
  Conference on Computer Science and Software Engineering}}
  \emph{(\bibinfo{series}{CASCON '19})}. \bibinfo{publisher}{IBM Corp.},
  \bibinfo{address}{USA}, \bibinfo{pages}{274--283}.
\newblock


\bibitem[Mastrangelo et~al\mbox{.}(2015)]%
        {Mastrangelo15use}
\bibfield{author}{\bibinfo{person}{Luis Mastrangelo}, \bibinfo{person}{Luca
  Ponzanelli}, \bibinfo{person}{Andrea Mocci}, \bibinfo{person}{Michele Lanza},
  \bibinfo{person}{Matthias Hauswirth}, {and} \bibinfo{person}{Nathaniel
  Nystrom}.} \bibinfo{year}{2015}\natexlab{}.
\newblock \showarticletitle{Use at Your Own Risk: The {J}ava {U}nsafe {API} in
  the Wild}. In \bibinfo{booktitle}{\emph{Proceedings of the 2015 ACM SIGPLAN
  International Conference on Object-Oriented Programming, Systems, Languages,
  and Applications}} \emph{(\bibinfo{series}{OOPSLA 2015})}.
  \bibinfo{publisher}{ACM}, \bibinfo{address}{New York, NY, USA},
  \bibinfo{pages}{695--710}.
\newblock
\showISBNx{9781450336895}
\urldef\tempurl%
\url{https://doi.org/10.1145/2814270.2814313}
\showDOI{\tempurl}


\bibitem[Munaiah et~al\mbox{.}(2017)]%
        {Munaiah17curating}
\bibfield{author}{\bibinfo{person}{Nuthan Munaiah}, \bibinfo{person}{Steven
  Kroh}, \bibinfo{person}{Craig Cabrey}, {and} \bibinfo{person}{Meiyappan
  Nagappan}.} \bibinfo{year}{2017}\natexlab{}.
\newblock \showarticletitle{Curating {G}it{H}ub for Engineered Software
  Projects}.
\newblock \bibinfo{journal}{\emph{Empirical Software Engineering}}
  \bibinfo{volume}{22}, \bibinfo{number}{6} (\bibinfo{date}{Dec.}
  \bibinfo{year}{2017}), \bibinfo{pages}{3219--3253}.
\newblock
\showISSN{1382-3256}
\urldef\tempurl%
\url{https://doi.org/10.1007/s10664-017-9512-6}
\showDOI{\tempurl}


\bibitem[Pothier and Tanter(2009)]%
        {Pothier09back}
\bibfield{author}{\bibinfo{person}{Guillaume Pothier} {and}
  \bibinfo{person}{\'Eric Tanter}.} \bibinfo{year}{2009}\natexlab{}.
\newblock \showarticletitle{Back to the Future: Omniscient Debugging}.
\newblock \bibinfo{journal}{\emph{IEEE Software}} \bibinfo{volume}{26},
  \bibinfo{number}{6} (\bibinfo{date}{Nov.} \bibinfo{year}{2009}),
  \bibinfo{pages}{78--85}.
\newblock
\urldef\tempurl%
\url{https://doi.org/10.1109/MS.2009.169}
\showDOI{\tempurl}


\bibitem[Rein et~al\mbox{.}(2018)]%
        {Rein18exploratory}
\bibfield{author}{\bibinfo{person}{Patrick Rein}, \bibinfo{person}{Stefan
  Ramson}, \bibinfo{person}{Jens Lincke}, \bibinfo{person}{Robert Hirschfeld},
  {and} \bibinfo{person}{Tobias Pape}.} \bibinfo{year}{2018}\natexlab{}.
\newblock \showarticletitle{Exploratory and Live, Programming and Coding: A
  Literature Study Comparing Perspectives on Liveness}.
\newblock \bibinfo{journal}{\emph{The Art, Science, and Engineering of
  Programming}} \bibinfo{volume}{3}, \bibinfo{number}{1}
  (\bibinfo{year}{2018}), \bibinfo{pages}{1:1--1:33}.
\newblock
\urldef\tempurl%
\url{https://doi.org/10.22152/programming-journal.org/2019/3/1}
\showDOI{\tempurl}


\bibitem[Reiss et~al\mbox{.}(2018)]%
        {Reiss18seede}
\bibfield{author}{\bibinfo{person}{Steven~P. Reiss}, \bibinfo{person}{Qi Xin},
  {and} \bibinfo{person}{Jeff Huang}.} \bibinfo{year}{2018}\natexlab{}.
\newblock \showarticletitle{S{EEDE}: Simultaneous Execution and Editing in a
  Development Environment}. In \bibinfo{booktitle}{\emph{Proceedings of the
  33rd ACM/IEEE International Conference on Automated Software Engineering}}
  \emph{(\bibinfo{series}{ASE 2018})}. \bibinfo{publisher}{ACM},
  \bibinfo{address}{New York, NY, USA}, \bibinfo{pages}{270--281}.
\newblock
\showISBNx{978-1-4503-5937-5}
\urldef\tempurl%
\url{https://doi.org/10.1145/3238147.3238182}
\showDOI{\tempurl}


\bibitem[Rocha et~al\mbox{.}(2019)]%
        {Rocha19compehending}
\bibfield{author}{\bibinfo{person}{Gilson Rocha}, \bibinfo{person}{Fernando
  Castor}, {and} \bibinfo{person}{Gustavo Pinto}.}
  \bibinfo{year}{2019}\natexlab{}.
\newblock \showarticletitle{Comprehending Energy Behaviors of Java {I/O}
  {API}s}. In \bibinfo{booktitle}{\emph{2019 ACM/IEEE International Symposium
  on Empirical Software Engineering and Measurement (ESEM)}}.
  \bibinfo{pages}{1--12}.
\newblock
\urldef\tempurl%
\url{https://doi.org/10.1109/ESEM.2019.8870158}
\showDOI{\tempurl}


\bibitem[Salda{\~n}a(2016)]%
        {Saldana16coding}
\bibfield{author}{\bibinfo{person}{Johnny Salda{\~n}a}.}
  \bibinfo{year}{2016}\natexlab{}.
\newblock \bibinfo{booktitle}{\emph{The Coding Manual for Qualitative
  Researchers} (\bibinfo{edition}{3rd} ed.)}.
\newblock \bibinfo{publisher}{SAGE Publishing}.
\newblock


\bibitem[Sul\'ir et~al\mbox{.}(2020)]%
        {Sulir20large}
\bibfield{author}{\bibinfo{person}{Mat\'u\v{s} Sul\'ir},
  \bibinfo{person}{Michaela Ba\v{c}\'ikov\'a}, \bibinfo{person}{Matej Madeja},
  \bibinfo{person}{Sergej Chodarev}, {and} \bibinfo{person}{J\'an Juh\'ar}.}
  \bibinfo{year}{2020}\natexlab{}.
\newblock \showarticletitle{Large-Scale Dataset of Local {J}ava Software Build
  Results}.
\newblock \bibinfo{journal}{\emph{Data}} \bibinfo{volume}{5},
  \bibinfo{number}{3} (\bibinfo{year}{2020}), \bibinfo{pages}{86}.
\newblock
\showISSN{2306-5729}
\urldef\tempurl%
\url{https://doi.org/10.3390/data5030086}
\showDOI{\tempurl}


\bibitem[Tan and Croft(2008)]%
        {Tan08empirical}
\bibfield{author}{\bibinfo{person}{Gang Tan} {and} \bibinfo{person}{Jason
  Croft}.} \bibinfo{year}{2008}\natexlab{}.
\newblock \showarticletitle{An Empirical Security Study of the Native Code in
  the {JDK}}. In \bibinfo{booktitle}{\emph{Proceedings of the 17th USENIX
  Security Symposium}} \emph{(\bibinfo{series}{SS'08})}.
  \bibinfo{publisher}{USENIX Association}, \bibinfo{address}{USA},
  \bibinfo{pages}{365--377}.
\newblock


\bibitem[Vall\'{e}e-Rai et~al\mbox{.}(2010)]%
        {Vallee-rai10soot}
\bibfield{author}{\bibinfo{person}{Raja Vall\'{e}e-Rai}, \bibinfo{person}{Phong
  Co}, \bibinfo{person}{Etienne Gagnon}, \bibinfo{person}{Laurie Hendren},
  \bibinfo{person}{Patrick Lam}, {and} \bibinfo{person}{Vijay Sundaresan}.}
  \bibinfo{year}{2010}\natexlab{}.
\newblock \showarticletitle{Soot: A Java Bytecode Optimization Framework}. In
  \bibinfo{booktitle}{\emph{CASCON First Decade High Impact Papers}}
  \emph{(\bibinfo{series}{CASCON '10})}. \bibinfo{publisher}{IBM Corp.},
  \bibinfo{address}{USA}, \bibinfo{pages}{214--224}.
\newblock
\urldef\tempurl%
\url{https://doi.org/10.1145/1925805.1925818}
\showDOI{\tempurl}


\end{thebibliography}

\end{document}